\documentstyle[aas2pp4]{article}
\newcommand{\rf}{\par\parshape 2 0mm 88mm 5mm
  83mm\noindent\footnotesize}

\begin{document}

\title{\vspace*{5em} Cosmological redshift distortion: deceleration, bias and \\ 
  density parameters from future redshift surveys of galaxies}

\author{Takahiro T. Nakamura$^1$, Takahiko Matsubara$^{1,2}$ \&
  Yasushi Suto$^{1,2}$} %
\affil{\altaffilmark{1} Department of Physics, University of Tokyo,
  Tokyo 113, Japan} %
\affil{\altaffilmark{2} Research Center for the Early Universe,
  University of Tokyo, Tokyo 113, Japan} %
\affil{\footnotesize nakamura@utaphp2.phys.s.u-tokyo.ac.jp, 
  matsu@phys.s.u-tokyo.ac.jp, suto@phys.s.u-tokyo.ac.jp}

%======================================================================
\begin{abstract}\noindent
  The observed two-point correlation functions of galaxies in redshift
  space become anisotropic due to the geometry of the universe as well
  as due to the presence of the peculiar velocity field. On the basis
  of linear perturbation theory, we expand the induced anisotropies of
  the correlation functions with respect to the redshift $z$, and
  obtain analytic formulae to infer the deceleration parameter $q_0$,
  the density parameter $\Omega_0$ and the derivative of the bias
  parameter $d\ln b/dz$ at $z=0$ in terms of the observable
  statistical quantities.  The present method does not require any
  assumption of the shape and amplitude of the underlying fluctuation
  spectrum, and thus can be applied to future redshift surveys of
  galaxies including the Sloan Digital Sky Survey.
  We also evaluate quantitatively the systematic error in estimating
  the value of $\beta_0 \equiv \Omega_0^{0.6}/b$ from a galaxy
  redshift survey on the basis of a conventional estimator for
  $\beta_0$ which neglects both the geometrical distortion effect and
  the time evolution of the parameter $\beta(z)$. If the magnitude
  limit of the survey is as faint as 18.5 (in B-band) as in the case
  of the Sloan Digital Sky Survey, the systematic error ranges between
  $-20\%$ and $10\%$ depending on the cosmological
  parameters. Although such systematic errors are smaller than the
  statistical errors in the current surveys, they will dominate the
  expected statistical error for future surveys.
\end{abstract}


\begin{center}
{\large\sl Accepted for publication in the Astrophysical Journal}
\end{center}

\unitlength=0.01\textwidth
\begin{picture}(100,0)(3,0)
  \put(0,110){\shortstack[l]{\sl UTAP: University of Tokyo,
      Theoretical Astrophysics \\ \sl RESCEU: Research Center for the
      Early Universe}} \put(90,110){\shortstack[l]{UTAP-262/97\\ 
      RESCEU-21/97}}
\end{picture}

\vspace{-10mm}
%======================================================================
\section{Introduction}\label{sec:1}

The volume size of the current galaxy redshift surveys is steadily
increasing. The Las Campanas redshift survey (Shectman et al. 1996),
for example, has reached a median redshift of $z\sim 0.1$ and the
number of galaxies is about 25000. In the near future, the Sloan
Digital Sky Survey (SDSS; e.g., Gunn \& Weinberg 1995) will complete a
spectroscopic survey of $\sim 10^6$ galaxies brighter than $18$
g-magnitude over $\pi$ steradians. The main purpose of this paper is
to stress that in interpreting observational results from these huge
surveys geometrical and evolutionary effects become important which
are intrinsic to non-zero redshifts since observations are performed
not on the constant time hypersurface but on the light cone.

It is supposed that global isotropy and homogeneity of the universe
guarantee the isotropy of the two-point correlation function of
galaxies in {\it real} space (i.e., the correlation function depends
only on the separation of galaxy pairs).  In reality, however, the
deviation from the pure Hubble flow or the peculiar velocity field
induces anisotropy in the observed correlation function in {\it
  redshift} space (Davis \& Peebles 1983; Kaiser 1987). This
redshift-space distortion of the correlation function has been
examined extensively in the literature both theoretically and
observationally in order to infer the value of $\beta_0 =
\Omega_0^{0.6} / b$ (Hamilton 1992; for a review Strauss \& Willick
1995), where $\Omega_0$ and $b$ are the density and bias parameters,
respectively.

If the redshift $z$ becomes substantially larger than $0$, the
geometry of the universe itself becomes an additional source for the
anisotropy in the two-point correlation functions (cosmological
redshift distortion): an intrinsically spherical object in real space
is elongated to an ellipse of axial ratio $H(z):z/S(z)$ along
line-of-sight in redshift-space (Alcock \& Paczynski 1979; Ryden
1995), where $H(z)$ is the Hubble parameter at $z$ and $S(z)$ is the
angular size distance (eqs.[\ref{5}] and [\ref{6}]). Ballinger,
Heavens, \& Peacock (1996) and Matsubara \& Suto (1996) developed an
idea to probe $\Omega_0$ and the cosmological constant, $\lambda_0$,
using this geometrical distortion in the correlation function having
quasar and/or galaxy clustering at high $z\gtrsim 1$ implicitly in
mind.  Two potential disadvantages of their methods are that they have
to assume a specific shape of the fluctuation spectra of the objects
before carrying out the analysis to find the best-fit values of
$\Omega_0$ and $\lambda_0$, and that the statistical errors due to the
limited number of high $z$ objects hamper the precise determination of
the cosmological parameters.

In this paper, we explore a possibility to use a sample of galaxy
redshift surveys at $z<1$. Although the geometrical distortion effect
becomes less important at $z<1$, the number of galaxies available for
the statistical analysis in a given redshift bin is larger by two
orders of magnitude than high $z$ quasars which compensates at least
partially the weak signal at $z<1$. In addition, we are successful in
evaluating the degree of the cosmological redshift distortion both
{\it analytically} and {\it independently} of the underlying
fluctuation spectrum shape by expanding the cosmological distortion
effect up to linear order in $z$ and using linear density
perturbation. Specifically we propose in \S\ref{sec:3} a method of
determining the deceleration parameter $q_0$ as well as $\Omega_0$ and
$d\ln b/dz|_{z=0}$ in this way, which would be possible if the
correlation function anisotropy can be accurately measured on large
scales where linear density perturbation theory applies. Our expansion
also enables us to estimate quantitatively how the value of $\beta_0$
determined from Hamilton's (1992) method systematically deviates from
its true value due to the neglect of the geometrical effect at $z\neq
0$. In \S\ref{sec:4} we show that systematic errors are smaller than
the statistical errors in the current surveys but that they should
dominate the expected statistical error for future surveys.  Section
\ref{sec:2} summarizes our notations which is extensively used
throughout the paper. Summary and conclusions are given in
\S\ref{sec:5}. We use the units in which $c$ and $H_0$ are unity.

%======================================================================
\section{Notations}\label{sec:2}

Consider two nearby galaxies located at a redshift of $z$ with a small
redshift separation $\Delta z (\ll z)$ and angular separation
$\Delta\theta$ on the sky. The separation between them, $s$, and the
direction cosine along line-of-sight, $\nu$, in the observable
(redshift-space) coordinates are
\begin{equation}\label{1}
  s := [ (\Delta z)^2 + (z\,\Delta\theta)^2 ]^{1/2} \,, 
\end{equation}
\begin{equation}
  \nu := \Delta z / s \,,
\end{equation}
while those in the comoving coordinates are
\begin{equation}\label{3}
  r := \{ [\Delta z/H(z)]^2 + [S(z) \Delta\theta]^2 \}^{1/2} \,,
\end{equation}
\begin{equation}\label{4}
  \mu := [\Delta z/H(z)]/ r \,,
\end{equation}
where 
\begin{equation}\label{5}
  H(z) := [ \Omega_0 (1+z)^3 - K(1+z)^2 + \lambda_0 ]^{1/2}
\end{equation}
is the Hubble parameter at $z$, $K:= \Omega_0 + \lambda_0 - 1$,
\begin{equation}\label{6}
  S(z) := \left\{
\begin{array}{ll}
  K^{-1/2} \sin(K^{1/2}\chi) & (K>0) \\ \chi & (K=0) \\ (-K)^{-1/2}
  \sinh((-K)^{1/2}\chi) & (K<0) \\ 
\end{array}\right.
\end{equation}
is the angular size distance (e.g., Peebles 1993) from us to $z$, and
\begin{equation}
  \chi(z) := \int_0^z dz'/ H(z') \,.
\end{equation}

The peculiar motions of galaxies induce anisotropy in the two-point
correlation function $\xi_{\rm s}(s,\nu;z)$ in redshift space (Kaiser
1987), while the real-space correlation function $\xi(r;z)$ is
isotropic (i.e., independent of the direction cosine $\mu$). Hamilton
(1992) expanded the redshift-space correlation function on the basis
of linear theory of density perturbation and distant-observer
approximation in terms of the Legendre polynomials with respect to the
direction cosine $\mu$ along line-of-sight at $z=0$ where $(s,\nu)$
and $(r,\mu)$ coincide. Matsubara \& Suto (1996) showed that
Hamilton's formula can be generalized to the $z \neq 0$ case even if
the cosmological distortion effect is taken into account:
\begin{equation}\label{8}
  \xi_{\rm s}(s,\nu;z) = \sum_{l=0}^2 \xi_{2l}(r;z) P_{2l}(\mu) \,,
\end{equation}
where $(s,\nu)$ and $(r,\mu)$ are related to each other by equations
(\ref{1}) to (\ref{4}), and the expansion coefficients are explicitly
given in terms of the real-space correlation function of galaxies
$\xi(r;z)$ as follows:
\begin{equation}\label{9}
  \xi_0(r;z) := [1 + \case23\beta(z) + \case15\beta^2(z) ] \xi(r;z)
  \,,
\end{equation}
\begin{equation}\label{10}
  \xi_2(r;z) := [ \case43\beta(z) + \case47\beta^2(z) ] \xi_{\rm
    d}(r;z) \,,
\end{equation}
\begin{equation}
  \xi_4(r;z) := \case8{35}\beta^2(z) [\xi_{\rm d}(r;z) +
  \case72\xi_{\rm q}(r;z) ] \,,
\end{equation}
\begin{equation}
  \xi_{\rm d}(r;z) := \xi(r;z) - 3\int_0^r \frac{dx}x (x/r)^3
  \xi(x;z) \,,
\end{equation}
\begin{equation}
  \xi_{\rm q}(r;z) := \int_0^r \frac{dx}x [3(x/r)^3 - 5(x/r)^5]
  \xi(x;z) \,.
\end{equation}
The $\beta$ parameter is defined as $\beta(z) := f(z)/b(z)$ with the
bias parameter $b$ and the following functions
\begin{eqnarray}
  \hspace{-2em} f(z) \!\!\!\!&:=&\!\!\!\! \frac{d\ln D}{d\ln a} =
  [(1+z)/H(z)]^2 / D(z) - [ 1+q(z) ] \\ \!\!\!\!&\simeq&\!\!\!\!
  \Omega^{0.6}(z) + \case1{70} \lambda(z) [ 1 + \case12 \Omega(z) ]
  \,,
\end{eqnarray}
\begin{equation}
  D(z) := H(z) \int_z^\infty dz' (1+z')/H^3(z') \,,
\end{equation}
\begin{equation}
  q(z) := -\frac{d\ln H}{d\ln a} - 1 = \case12\Omega(z) - \lambda(z)
  \,,
\end{equation}
\begin{equation}
  \Omega(z) := \Omega_0(1+z)^3 / H^2(z) \,,\quad \lambda(z) :=
  \lambda_0 / H^2(z)
\end{equation}
(e.g., Peebles 1980; Lahav et al. 1991). Defining
\begin{equation}\label{19}
  X(r;z) := [ \xi_0 (r;z) - 3\int_0^1 dx \, x^2 \xi_0(rx;z) ]/
  \xi_2(r;z)
\end{equation}
(Hamilton 1992), one finds from equations (\ref{9}) and (\ref{10})
that
\begin{equation}\label{20}
  \beta(z) = \case32 / [ X - \case12 + (X^2 + \case27 X -
  \case15)^{1/2} ] \,.
\end{equation}

%======================================================================
\section{Measuring the deceleration parameter from redshift-distortion
  at small redshifts}
\label{sec:3}

Matsubara \& Suto (1996) showed that the shape of contour curves of
$\xi_{\rm s}(s,\nu;z)$ at $z\gtrsim 1$ sensitively depends on
$\Omega_0$ and $\lambda_0$, and proposed to test the non-vanishing
$\lambda_0$ in particular by the comparison of theoretical prediction
and the observation.  As mentioned in \S\ref{sec:1}, however, this
method is model-dependent in the sense that such a determination of
$\Omega_0$ and $\lambda_0$ is possible only when one assumes a priori
a specific power spectrum, such as that of the cold dark matter (CDM)
model, in order to fix the (unobservable) real-space correlation
function $\xi(r;z)$.  This is because the comoving coordinates
$(r,\mu)$ themselves depend on $\Omega_0$ and $\lambda_0$ at $z\neq 0$
(eqs.[\ref{3}] and [\ref{4}]): one cannot extract the multipole
components $\xi_{2l}$ directly from the observed correlation function
$\xi_{\rm s}$ (eq.[\ref{8}]) without a priori knowledge of $\Omega_0$
and $\lambda_0$, but rather the observed $\xi_{\rm s}$ need be
compared with the one calculated theoretically from equation (\ref{8})
with an assumed model power spectrum. Thus $\xi_{2l}(r;z)$ are not
directly observable quantities at $z\neq 0$, but the observable ones
should be expressed in terms of the redshift-space variables like:
\begin{equation}\label{21}
  \zeta_{2l}(s;z) := (2l + \case12) \int_{-1}^1 d\nu\, P_{2l}(\nu)
  \, \xi_{\rm s} (s,\nu;z) \,,
\end{equation}
i.e., the multipole expansion with respect to the observable
coordinate $\nu$ instead of $\mu$ [note that $\zeta_{2l}(s;0) =
\xi_{2l} (r;0)$].

In this sense, it is desirable to rewrite equation (\ref{8}) entirely
in terms of $\zeta_{2l}(s;z)$ and $P_{2l}(\nu)$, but this expansion
has infinite numbers of terms. However we find that this expansion
becomes finite if all the variables are expanded up to linear order in
$z$. This perturbation analysis is quite relevant since we have
specifically in mind the SDSS galaxy redshift survey ($z\lesssim 0.2$)
throughout the present paper.  With this prescription, we obtain
explicit expressions for the deceleration parameter $q_0$, the density
parameter $\Omega_0$ and the derivative of the bias parameter $d\ln
b/dz$ at $z=0$ in terms of the observable quantities.  As shown below,
the advantage of the present method is that the resulting formulae are
written only in terms of the observables and independent of the model
power spectrum unlike the high $z$ analysis (Matsubara \& Suto 1996).

To linear order in $z$, $(r,\mu)$ and $(s,\nu)$ are written from
equations (\ref{1}) to (\ref{4}) as
\begin{equation}
  r \simeq [ 1 - \case12 (1+\nu^2) (1+q_0) z ] s \,,
\end{equation}
\begin{equation}
  \mu \simeq [ 1 - \case12 (1-\nu^2) (1+q_0) z ] \nu \,,
\end{equation}
where and hereafter we use the symbol $\simeq$ to indicate that the
equality is valid only up to linear order in $z$. Substituting these
into equation (\ref{8}) and using equation (\ref{21}), we obtain
\begin{equation}
  \xi_{\rm s} (s,\nu;z) \simeq \sum_{l=0}^3 \zeta_{2l}(s;z) \,
  P_{2l}(\nu) \,,
\end{equation}
\begin{eqnarray}\label{25}
  \zeta_0(s;z) \!\!\!\!&\simeq& \nonumber \\ && \hspace{-5em}
  \xi_0(s;z) - \case23 (1 + \case45\beta_0 + \case9{35}\beta_0^2)
  \frac{\partial \xi(s;0)}{\partial\ln s} (1+q_0)z \,,
\end{eqnarray}
\begin{eqnarray}\label{26}
  \zeta_2(s;z) \!\!\!\!&\simeq&\!\!\!\! \xi_2(s;z) + [
  (\case{20}7\beta_0 + \case43\beta_0^2) \xi_{\rm d}(s;0) \nonumber \\ 
  && \hspace{-2em} - \case13 (1 + \case{26}7\beta_0 +
  \case{11}7\beta_0^2) \frac{\partial\xi(s;0)} {\partial\ln s} ]
  (1+q_0) z \,,
\end{eqnarray}
\begin{eqnarray}\label{27}
  \zeta_4(s;z) \!\!\!\!&\simeq&\!\!\!\! \xi_4(s;z) + [ (\case87\beta_0
  + \case{24}{11}\beta_0^2) \xi_{\rm d}(s;0) + \case{32}{11}\beta_0^2
  \xi_{\rm q}(s;0) \nonumber \\ && - \case8{35}(\beta_0
  + \case{13}{11}\beta_0^2) \frac{\partial\xi(s;0)}{\partial\ln s} ]
  (1+q_0) z \,,
\end{eqnarray}
\begin{eqnarray}\label{28}
  \zeta_6(s;z) \!\!\!\!&\simeq& \nonumber \\ && \hspace{-5em}
  \case4{33} \beta_0^2 [ 4\xi_{\rm d}(s;0) + 9 \xi_{\rm q}(s;0) -
  \case27 \frac{ \partial\xi(s;0) } { \partial\ln s }] (1+q_0) z \,,
\end{eqnarray}
where $\beta_0 := \beta(0)$. Equations (\ref{25}) to (\ref{28})
represent the geometrical effect on the redshift-space distortion of
the correlation function to linear order in $z$. We also calculate its
time evolution for consistency:
\begin{eqnarray}\label{29}
  \xi_0(s;z) \!\!\!\!&\simeq& \nonumber \\ && \hspace{-5em}
  \xi_0(s;0) - z [ (1 + \case13\beta_0)\phi_0 + (\case13\beta_0 +
  \case15\beta_0^2 ) \psi_0 ] \xi(s;0) \,,
\end{eqnarray}
\begin{equation}\label{30}
  \xi_2(s;z) \simeq \xi_2(s;0) - z[ \case23\beta_0\phi_0 +
  (\case23\beta_0 + \case47\beta_0^2 ) \psi_0] \xi_{\rm d}(s;0) \,,
\end{equation}
\begin{equation}\label{31}
  \xi_4(s;z) \simeq (1 - \psi_0 z)\xi_4(s;0) \,,
\end{equation}
where 
\begin{equation}\label{32}
  \phi_0 := - \left. \frac{d}{dz} \ln|\xi| \right|_{z=0} = 2 f_0 - 2
  \left. \frac{d}{dz} \ln b \right|_{z=0} \,,
\end{equation}
\begin{equation}\label{33}
  \psi_0 := - \left. \frac{d}{dz} \ln |\beta^2\xi| \right|_{z=0} = 3
  \Omega_0/f_0 - 2(1 - q_0)
\end{equation}
(the subscript 0 denotes variables at the present epoch $z=0$). In the
second equality of equations (\ref{32}) and (\ref{33}), it is assumed
that the real-space correlation function $\xi(r;z)$ evolves as
$\propto [b(z) D(z)]^2$ with $z$. Note that $\psi_0$ measures the time
evolution of the velocity correlation.

Next we integrate the observable quantities $[ \xi_{2l}(s;0) -
\zeta_{2l}(s;z) ] /z$ over $s$ with some appropriate weights, and
eliminate the derivative terms $\partial \xi/ \partial \ln s$ in
equations (\ref{25}) to (\ref{28}). Performing three such integrals,
we obtain linear equations of the form:
\begin{equation}\label{34}
  \left[
\begin{array}{cccc}
  c_{11} & c_{12} & c_{13} \\ c_{21} & c_{22} & c_{23} \\ c_{31} &
  c_{32} & c_{33} \\
\end{array}\right]
\left[
\begin{array}{c}
  1+q_0 \\ \phi_0 \\ \psi_0
\end{array}\right]
\simeq \left[
\begin{array}{c}
  d_1 \\ d_2 \\ d_3
\end{array}\right] \,,
\end{equation}
where $d_i$'s are the integrals of $[\xi_{2l}(s;0) - \zeta_{2l}(s;z)]
/ z$ (an explicit set of examples is given by eqs.[\ref{35}] to
[\ref{37}]), and $c_{ij}$'s consist of $\beta_0$ and $\xi(r;0)$. If we
assume that $c_{ij}$'s are known from observation at $z\sim 0$, then
we can infer the values of $1+q_0$, $\phi_0$ and $\psi_0$ by solving
equation (\ref{34}) with respect to them. In this way the cosmological
parameters as well as (local) evolution of the bias can be determined
in principle, without assuming any specific cosmological models such
as CDM, by observing the multipole components $\zeta_{2l}(s;z)$ at
small $z$.

The sixth moment $\zeta_6$, which does not appear in Hamilton's
formula at $z=0$, shows up in our expansion in redshift space at $z
\neq 0$ (within the linear density perturbation) and, according to
equation (\ref{25}), the solution for $1+q_0$ has rather simple form
in terms of an integral of $\zeta_6$. However, higher-order moments
would become progressively difficult to be determined reliably, and
also are likely to be contaminated by non-linear effects (Cole, Fisher
\& Weinberg 1994, 1995). Therefore we eliminate the use of the
sixth-order moment, and find that the following choice of $d_i$'s
leads to the simplest form of the solution for $1+q_0$ among what we
have examined:
\begin{equation}\label{35}
  d_1(s;z) = \int_0^1 dx (3x^2 - 5x^4 ) [ \xi_0(sx;0) - \zeta_0(sx;z)
  ]/z \,,
\end{equation}
\begin{equation}
  d_2(s;z) = 3 \int_0^1 dx \, x^4 [ \xi_2(sx;0) - \zeta_2(sx;z) ]/z
  \,,
\end{equation}
\begin{equation}\label{37}
  d_3(s;z) = \frac{35}4 \int_0^1 dx \, x [ \xi_4(s/x;0) -
  \zeta_4(s/x;z) ]/z \,.
\end{equation}
Then it follows from equations (\ref{25}) to (\ref{31}) that
\begin{equation}\label{38}
  c_{11}(s) = -\case23(1 + \case45\beta_0 + \case9{35}\beta_0^2) [
  2\xi_{\rm d}(s;0) + 5\xi_{\rm q}(s;0) ]
\end{equation}
\begin{equation}
  c_{12}(s) = (1 + \case13\beta_0) \xi_{\rm q}(s;0) \,,
\end{equation}
\begin{equation}
  c_{13}(s) = (\case13\beta_0 + \case15\beta_0^2) \xi_{\rm q}(s;0) \,,
\end{equation}
\begin{eqnarray}
  c_{21}(s) \!\!\!\!&=&\!\!\!\! (1 + \case{26}7\beta_0 +
  \case{11}7\beta_0^2) \xi_{\rm d}(s;0) \nonumber \\ && + (1 +
  8\beta_0 + \case{25}7\beta_0^2) \xi_{\rm q}(s;0) \,,
\end{eqnarray}
\begin{equation}
  c_{22}(s) = -\beta_0 \xi_{\rm q}(s;0) \,,
\end{equation}
\begin{equation}
  c_{23}(s) = -(\beta_0 + \case67\beta_0^2) \xi_{\rm q}(s;0) \,,
\end{equation}
\begin{equation}
  c_{31}(s) = -2(\beta_0 + \case{13}{11}\beta_0^2) \xi_{\rm d}(s;0) -
  \case{40}{11} \beta_0^2 \xi_{\rm q}(s;0) \,,
\end{equation}
\begin{equation}
  c_{32} = 0 \,,
\end{equation}
\begin{equation}\label{46}
  c_{33}(s) = \beta_0^2 \xi_{\rm q}(s;0) \,.
\end{equation}
From equations (\ref{38}) to (\ref{46}), we solve equation (\ref{34})
for $1+q_0$ as
\begin{equation}\label{47}
  1+q_0 \simeq \beta_0^2 \frac{d_1}{\Delta} + (\beta_0 +
  \case13\beta_0^2) \frac{d_2}{\Delta} + (1 + \case67\beta_0 +
  \case3{35}\beta_0^2) \frac{d_3}{\Delta} \,,
\end{equation}
where
\begin{equation}\label{48}
  \Delta(s) = (\beta_0 + \case{15}{11}\beta_0^2 +
  \case5{11}\beta_0^3 + \case5{231}\beta_0^4 )[ \xi_{\rm q}(s;0) -
  \xi_{\rm d}(s;0) ] \,,
\end{equation}
and $d_1$, $d_2$ and $d_3$ are defined in equations (\ref{35}) to
(\ref{37}). The solutions for $\phi_0$ and $\psi_0$ (eqs.[\ref{32}]
and [\ref{33}]) are presented in Appendix \ref{sec:a}.

\begin{figure}[tbh]
\centerline{\epsfxsize=0.95\hsize {\epsfbox{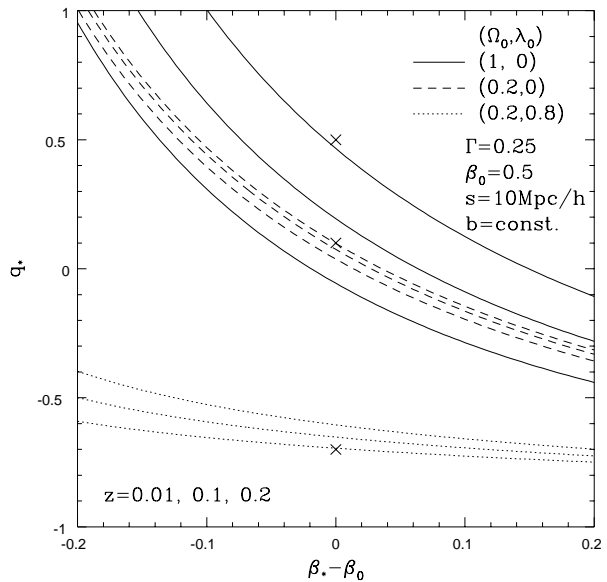}} }
\caption{Inferred value $q_*$ of the deceleration parameter $q_0$
  from equation (\protect\ref{47}), versus $\beta_*$ which is
  substituted for $\beta_0 = \Omega_0^{0.6}/b$ in the right hand sides
  of equations (\protect\ref{47}) and (\protect\ref{48}). The crosses
  indicate the location of the true values of $q_0=\frac12\Omega_0 -
  \lambda_0$. The three curves for each model correspond to the
  $z=0.01$, $=0.1$ and $0.2$ cases, with smaller $z$ corresponding to
  the closer curve to the crosses.\label{fig1}}
\end{figure}

Figure \ref{fig1} illustrates the extent to which equation (\ref{47})
is useful in determining $q_0$, where we plot the inferred value $q_*$
of $q_0$ (left hand side of eq.[\ref{47}]) versus $\beta_*$ which is
substituted for $\beta_0$ in the right hand sides of equations
(\ref{47}) and (\ref{48}) for three representative sets of $\Omega_0$
and $\lambda_0$ assuming $\beta_0=0.5$ in all cases. In the figure we
calculate $\xi_{2l}$ and $\zeta_{2l}$ from equations (\ref{8}) and
(\ref{21}) assuming that $b$ is constant and $\xi(r;z) \propto
D^2(z)$. The integrals in equations (\ref{35}) to (\ref{37}) are
performed numerically over $0.01 < x < 1 $ in logarithmic
interval. The real-space correlation function $\xi(r;0)$ is calculated
from the scale-invariant ($n=1$) CDM-like power spectrum in Bardeen et
al. (1986) with the shape parameter ${\mit\Gamma} = 0.25$ irrespective
of $\Omega_0$ and $\lambda_0$, since the spectrum with these
parameters fits the observed linear power spectrum fairly well
(Peacock \& Dodds 1994).  The crosses show the true value of $q_0 :=
\frac12\Omega_0 - \lambda_0$ for each $\Omega_0$ and $\lambda_0$.  The
three curves for each model correspond to the $z=0.01$, $=0.1$ and
$0.2$ cases, with smaller $z$ corresponding to the the closer curve to
the crosses.

To determine $q_0$ in practice, one would probably bin the
observational data in $z$ with the width of $\Delta z\ll z$, calculate
the multipole components of the correlation function anisotropy within
each bin, and substitute them for $\xi_{2l}(s;0)$ and $\zeta_{2l}
(s;z)$ in equations (\ref{35}) to (\ref{37}). Since these equations
contain subtractions of two similar quantities, one needs very
accurate data of them to avoid roundoff errors. Thus the actual curve
drawn from observation may have very large error bars, so a sample as
large as the SDSS catalogue is required for that purpose. Also one has
to guess the trial value $\beta_*$ of $\beta_0$ from observation at
$z\sim 0$.

The curves in the figure do not pass exactly through the crosses since
we have neglected terms of order $z^2$, so the deviations of the
curves from the crosses represent the second order
contributions. Alternatively, by extrapolating the curves to $z\sim
0$, one may infer a curve on which the true $q_0$ lies.  The
$z$-dependence is very large for the Einstein-de Sitter ($\Omega_0=1$
and $\lambda_0=0$) model and one may not be able to distinguish the
Einstein-de Sitter and open ($\Omega_0=0.2$ and $\lambda_0=0$) models.
We expect, however, that $\lambda_0 = 0$ and $\lambda_0\sim 1$ models
can be distinguished clearly. Moreover, unlike the Einstein-de Sitter
and open models, the curve in the $\Lambda$ model ($\Omega_0 = 0.2$
and $\lambda_0 = 0.8$) is quite insensitive to the value of $\beta_*$
implying that $q_0$ may be inferred very reliably even if the
uncertainty in the estimate of $\beta_0$, i.e., $\beta_*-\beta_0$, is
fairly large.

\begin{figure}[tbh]
\centerline{\epsfxsize=0.95\hsize {\epsfbox{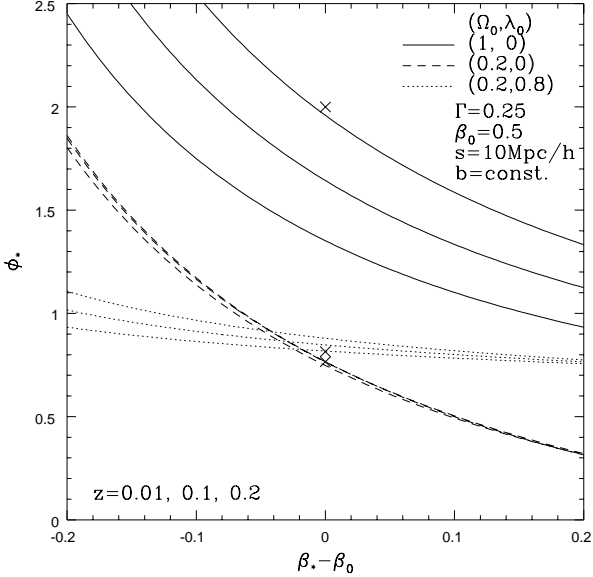}} }
\caption{Same as Fig.\protect\ref{fig1} on $\phi_0 =
  2\Omega_0^{0.6} - 2d\ln b/dz|_{z=0}$ (see eqs.[\protect\ref{32}] and
  [\protect\ref{a1}]).\label{fig2}}
\end{figure}

\begin{figure}[tbh]
\centerline{\epsfxsize=0.95\hsize {\epsfbox{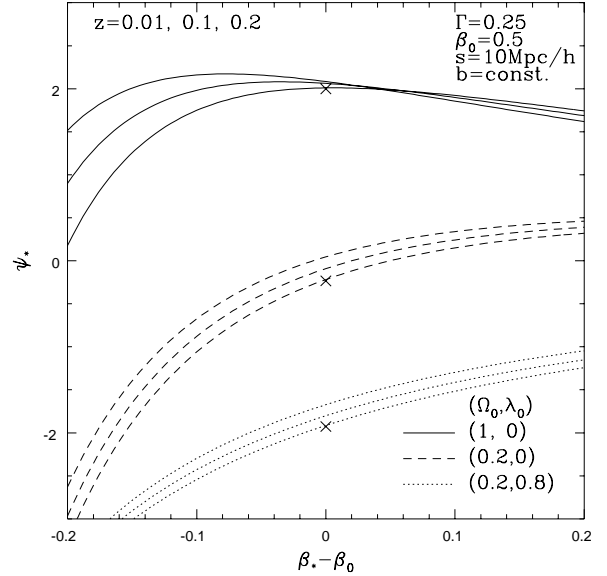}} }
\caption{Same as Fig.\protect\ref{fig1} on $\psi_0 =
  3\Omega_0^{0.4} - 2(1-q_0)$ (see eqs.[\protect\ref{33}] and
  [\protect\ref{a2}]).\label{fig3}}
\end{figure}

Figures \ref{fig2} and \ref{fig3} show the same plots for $\phi_0$ and
$\psi_0$ (eqs.[\ref{32}] and [\ref{33}]) as in Fig.\ref{fig1}, using
equations (\ref{a3}) to (\ref{a8}). We expect that $\phi_*$ can
distinguish high-$\Omega_0$ and low-$\Omega_0$ models (if the bias
$b(z)$ does not evolve), and $\psi_*$ can distinguish all the three
models. Thus one can put strong constraints on $\Omega_0$, $\lambda_0$
and $\beta_0$ by combining the three inferred quantities $q_*$,
$\phi_*$ and $\psi_*$. If one determines these parameters with the
above methods, then $\Omega_0$, $\lambda_0$ and the evolution of bias
$d\ln b/dz|_{z=0}$ can also be determined in principle, since these
have a one-to-one correspondence with $q_0$, $\phi_0$ and $\psi_0$.

%======================================================================
\section{Geometrical and evolutionary effects on the estimates of
  $\beta_0$}\label{sec:4}

As noted in the last section, the directly observable quantities are
not the multipole components $\xi_{2l}$ but $\zeta_{2l}$.  Thus
equation (\ref{20}) is useful only at $z\sim 0$ to infer the value of
$\beta_0 := \beta(0)$. In addition to the difference between $\mu$ and
$\nu$, the time-evolution of both the peculiar velocity field and the
bias also affects the estimates of $\beta_0$. Therefore, as the depth
of the redshift survey increases, the value of $\beta_0$ estimated
from equation (\ref{20}) in a straightforward manner would deviate
systematically from the true value.

The systematic deviation $\Delta\beta$ as a function of $z$ can be
evaluated by substituting the observable $\zeta_{2l}$ with $\xi_{2l}$
in equation (\ref{19}):
\begin{equation}\label{49}
  Z(s;z) := [ \zeta_0(s;z) - 3\int_0^1 dx \, x^2 \zeta_0(sx;z) ]
  / \zeta_2(s;z) \,.
\end{equation}
The resulting estimator of the $\beta$-parameter according to
equation (\ref{20}) is
\begin{eqnarray}
  \beta_{\rm obs}(z) \!\!\!\!&=&\!\!\!\! \case32 / [ Z - \case12 +
  (Z^2 + \case27 Z - \case15)^{1/2} ] \nonumber \\ &=&\!\!\!\! \beta_0
  + \Delta\beta_{\rm evol}(z) + \Delta\beta_{\rm geom}(z) \,,
\end{eqnarray}
where we define
\begin{equation}\label{51}
  \Delta\beta_{\rm evol}(z) := \beta(z) - \beta_0 \,,
\end{equation}
\begin{equation}\label{52}
  \Delta\beta_{\rm geom}(z) := \beta_{\rm obs}(z) - \beta(z) \,.
\end{equation}
The former correction term $\Delta\beta_{\rm evol}$ simply represents
the time-evolution of both the velocity field and the bias, while the
latter term arises from the geometrical effect which we have
discussed. Using equations (\ref{25}) to (\ref{31}), these are
expressed up to linear order in $z$ as
\begin{equation}\label{53}
  \Delta\beta_{\rm evol}(z) \simeq \left.{d\beta \over
      dz}\right|_{z=0} z = \case12\beta_0 (\phi_0-\psi_0) z
  \,,
\end{equation}
\begin{eqnarray}\label{54}
  \Delta\beta_{\rm geom}(z) \!\!\!\!&\simeq&\!\!\!\! - (1+q_0)z / ( 1
  + \case67 \beta_0 + \case3{35} \beta_0^2 ) \nonumber \\ &&
  \hspace{-6em} \times [ \case14 (1 + \case{12}7 \beta_0 +
  \case{34}{35} \beta_0^2 + \case4{21}\beta_0^3 + \case1{49} \beta_0^4
  ) ( 3 + \frac{ \partial\ln|\xi_{\rm d}| }{ \partial\ln s } |_{z=0} )
  \nonumber \\ && \hspace{-4em} - \case17 (\beta_0 - \case15\beta_0^2 -
  \case{11}{35} \beta_0^3 - \case17\beta_0^4) ] \,.
\end{eqnarray}

\begin{figure}[tbh]
\centerline{\epsfxsize=0.95\hsize {\epsfbox{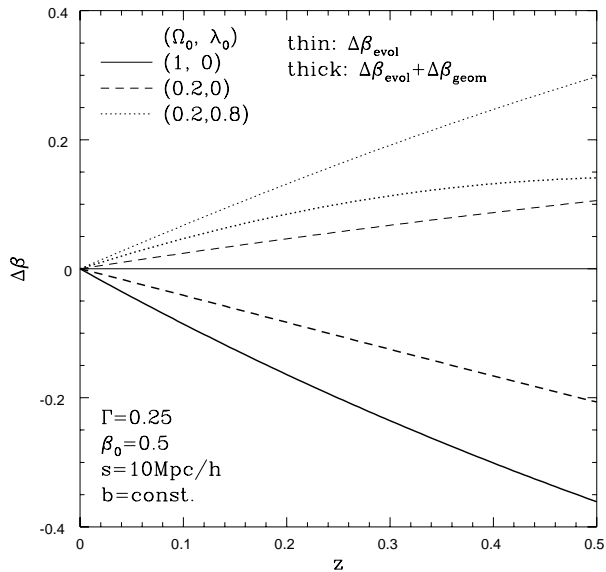}} }
\caption{Systematic deviation $\Delta\beta$ in the estimates of
  $\beta_0=\Omega_0^{0.6}/b$ versus redshift. Thin and thick curves
  correspond to $\Delta\beta_{\rm evol}$ (eq.[\protect\ref{51}]) and
  $\Delta\beta_{\rm evol} + \Delta\beta_{\rm geom}$
  (eq.[\protect\ref{52}]), respectively.\label{fig4}}
\end{figure}

Figure \ref{fig4} plots $\Delta\beta(z)$ for three sets of values
$(\Omega_0,\lambda_0)$, again assuming $\beta_0 = 0.5$. Thin and thick
curves correspond to $\Delta\beta_{\rm evol}$ (eq.[\ref{51}]) and
$\Delta\beta_{\rm evol} + \Delta\beta_{\rm geom}$ (eq.[\ref{52}]),
respectively. In the figure we set the bias parameter $b$ to be
constant with $z$ so $\beta(z)\propto f(z)$.  As in Figures \ref{fig1}
to \ref{fig3}, $\zeta_{2l}(s)$ were calculated from equations
(\ref{8}) and (\ref{21}), using the CDM-like power spectrum of
${\mit\Gamma} = 0.25$ and $n = 1$ (see \S\ref{sec:3}).  The integral
in equation (\ref{49}) was performed numerically over $0.01 < x < 1$
using bins with the logarithmically equal interval.

As seen from Fig.\ref{fig4} and equations (\ref{53}) and (\ref{54}),
$\Delta\beta_{\rm evol}$ and $\Delta\beta_{\rm geom}$ are generally
positive and negative, respectively; for $\lambda_0 = 0$ models, the
geometrical effect (proportional to $1+q_0$) is important and
dominates the evolutionary effect so that the total $\Delta\beta$ is
negative. On the contrary, the opposite is true for $\lambda_0\sim 1$
models due to the steady character of the de Sitter space-time, so
$\Delta\beta$ is positive. In all the models in Figure \ref{fig4}, the
systematic deviation $\Delta\beta$ amounts to more than 10\% beyond
the redshift of $0.1$.

To be more realistic, the multipole components obtained from actual
redshift survey (whose limiting magnitude is $m$) may be written as an
average over the redshift:
\begin{equation}\label{55}
  \langle \zeta_{2l} (s;m) \rangle = \int_0^\infty dz \, w(z) \,
  \zeta_{2l} (s;z) \,.
\end{equation}
The weight function $w(z)$, with the normalization $\int_0^\infty dz
\, w(z) = 1$, is contributed from the survey volume and the selection
for galaxy brightness:
\begin{equation}
  w(z) \propto [S^2(z)/H(z)] \,\int_{L(z)}^\infty \Phi(L') dL' \,,
\end{equation}
\begin{equation}
  L(z) = 4\pi [(1+z)\,S(z)]^2 f_{\min} \,,
\end{equation}
where $\Phi(L)dL$ is the galaxy luminosity function and $f_{\min}$ is
the faintest flux corresponding to the limiting magnitude
$m$. Substituting $\langle \zeta_{2l} (s;m) \rangle$ instead of
$\zeta_{2l} (s;z)$ in equation (\ref{49}), we can estimate the
systematic deviation $\Delta\beta(m) : = \beta_{\rm obs}(m) - \beta_0$
in the estimates of $\beta_0$.

\begin{figure}[tbh]
\centerline{\epsfxsize=0.95\hsize {\epsfbox{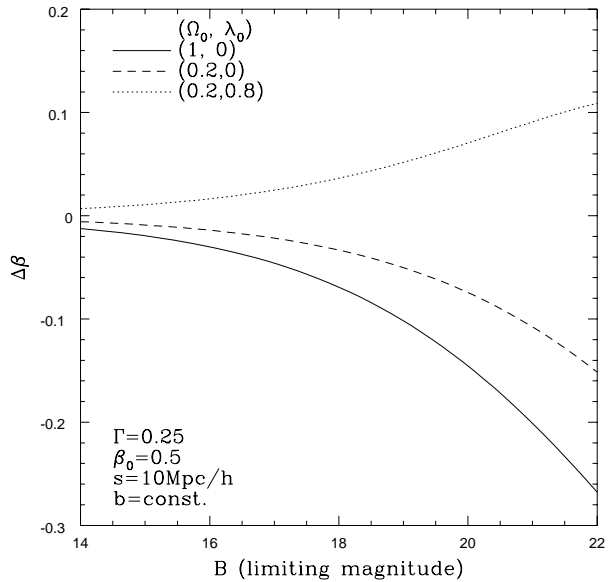}} }
\caption{Systematic deviation $\Delta\beta$ in the
  estimates of $\beta_0=\Omega_0^{0.6}/b$ versus the limiting
  magnitude of redshift surveys in B band.\label{fig5}}
\end{figure}

Figure \ref{fig5} plots $\Delta\beta$ versus the limiting magnitude of
surveys in B band. We adopt the parameters for the galaxy luminosity
function from Loveday et al. (1992), and use the same cosmological
parameters as in Fig.\ref{fig4}. Incidentally, we perform the same
calculation except that the correlation function is constant with $z$,
i.e., $\xi(r;z) = \xi(r;0)$, and find that little difference is shown
up compared with Fig.\ref{fig5}. Thus the results in Fig.\ref{fig5} do
not depend on how the density contrast evolves with time.

The Durham/UKST redshift survey (Ratcliffe et al. 1996) and the
Stromlo-APM redshift survey (Loveday et al.  1996) estimate that
$\beta_0 = 0.55\pm 0.12$ and $0.48\pm 0.12$, respectively, from $\sim
10^3$ galaxies of $B\lesssim 17$ mag. From Fig.\ref{fig5}, the
systematic deviation $\Delta\beta$ is less than $\sim 0.05$ (10\%) for
$B<17$, which is smaller than the above statistical errors $\pm 0.12$.
On the other hand, future redshift surveys of galaxies, the SDSS, for
example, will observe $\sim 10^6$ galaxies with $B\lesssim 18.5$
(Fukugita, Shimasaku \& Ichikawa 1995).  This huge number of galaxies
should greatly reduce the statistical error in the estimates of
$\beta_0$ and the systematic deviation pointed out in this paper
should dominate the statistical errors.

%======================================================================
\section{Summary and conclusion}\label{sec:5}

In this paper we have shown that, when interpreting observational
results from forthcoming huge redshift surveys, it is necessary to
consider geometrical and evolutionary effects which are intrinsic to
non-zero redshifts, due to the fact that observations are performed on
the light cone. With the geometrical and light-cone effects,
Hamilton's (1992) method of determining $\beta_0$ does not work in a
straightforward manner. This is why Matsubara \& Suto (1996) proposed
the model-dependent comparison with observation. In fact we are
working on a method to determine cosmological parameters independently
of the underlying fluctuation spectrum (Matsubara et al. in
preparation). For $z \ll 1$ as we consider in the present paper,
however, the straightforward application of the Hamilton method works
in determining the parameters independently of the spectra as we
showed explicitly.

In \S\ref{sec:3}, we have explicitly obtained the expressions for the
observable multipole components of the redshift-space correlation
function up to linear order in the redshift, bearing specifically the
redshift survey data of $z < 1$ like the SDSS in mind. Then we derived
analytic formulae, in linear theory and the distant observer
approximation, to infer the cosmological parameters $q_0$, $\phi_0$
and $\psi_0$ (eqs.[\ref{32}] and [\ref{33}] for definition) in terms
of these multipole components at small $z$. Assuming that $\beta_0 =
\Omega_0^{0.6}/b$ and $\xi(r;0)$ are independently given from
observations at $z\sim 0$, our formulae (eqs.[\ref{47}], [\ref{a1}]
and [\ref{a2}]) provide model-independent determination of these
parameters from observation of the redshift-space correlation function
at small $z$. To determine them in practice, one would probably bin
the observational data in $z$ with the width of $\Delta z\ll z$,
calculate the multipole components $\zeta_{2l}$ of the correlation
function anisotropy within each bin, and substitute them in equations
(\ref{35}) to (\ref{37}). Since these equations contain subtractions
of two similar quantities, one needs very accurate data of them to
avoid roundoff errors, so a sample as large as the SDSS catalogue is
required for that purpose. Also one has to guess the value of
$\beta_0$ in the formulae from observation at $z\sim 0$. If one
determines $q_0$, $\phi_0$ and $\psi_0$ in this way, then $\Omega_0$,
$\lambda_0$ and the evolution of bias $d\ln b/dz|_{z=0}$ are also
obtained.

In \S\ref{sec:4}, we pointed out that the value of $\beta_0$ estimated
by existing methods deviates systematically from its true value, due
to the time-evolution of both the velocity field and the bias, and due
also to the cosmological geometry.  We quantitatively estimated the
systematic deviation $\Delta\beta$ when Hamilton's formula
(eq.[\ref{20}]) is applied to data of large-scale redshift surveys,
and found that $\Delta\beta/\beta_0 \lesssim 10\%$ for $B<17$
(Fig.\ref{fig1}) which is smaller than the statistical errors of
currently estimated $\beta_0$ from existing surveys (Ratcliffe et
al. 1996; Loveday et al. 1996). For the SDSS ($B < 18.5$; Fukugita et
al. 1995) in which the statistical error is expected to be
significantly reduced, we found $\Delta\beta/\beta_0$ ranges between
$-20\%$ and $10\%$ depending on the cosmological parameters. Thus the
systematic deviation is likely to dominate the statistical error in
the next-generation, deep and wide redshift surveys.

The remaining future work is to examine, using a mock sample from an
$N$-body simulation, whether our formulae can practically be applied
to determine the cosmological parameters. We plan to ``observe'' the
redshift dependence of the multipole components $\zeta_{2l}$ in a mock
sample like the SDSS, binning the $N$-body data in $z$ (Magira et al.
in preparation). At the same time, it is desired to work out more
efficient and accurate ways of obtaining the multipole components from
survey or mock data on large scales where linear theory applies,
and/or to include non-linear effects in our analytic calculations.

\bigskip

We thank the referee, David Weinberg, for several useful comments on
the earlier manuscript, which helped improve the presentation of the
paper.  T.T.N. gratefully acknowledges support from a JSPS (Japan
Society of Promotion of Science) fellowship.  This work is supported
in part by grants-in-aid by the Ministry of Education, Science, Sports
and Culture of Japan (4125, 07CE2002, 07740183).

%======================================================================
\appendix
\section*{Appendix}
\section{Solutions for $\phi_0$ and $\psi_0$}\label{sec:a}

In terms of the integrals $d_1$, $d_2$ and $d_3$ (eqs.[\ref{35}] to
[\ref{37}]), the solutions of equation (\ref{34}) for $\phi_0$
(eq.[\ref{32}]) and $\psi_0$ (eq.[\ref{33}]) are written as
\begin{equation}\label{a1}
  \phi_0 \simeq (\Delta_{\phi1} d_1 + \Delta_{\phi2} d_2 +
  \Delta_{\phi3} d_3 ) / \Delta \,,
\end{equation}
\begin{equation}\label{a2}
  \psi_0 \simeq (\Delta_{\psi1} d_1 + \Delta_{\psi2} d_2 +
  \Delta_{\psi3} d_3 ) / \Delta \,,
\end{equation}
where $\Delta$ is defined in equation (\ref{48}), and
\begin{eqnarray}\label{a3}
  \Delta_{\phi1}(s) \!\!\!\!&=&\!\!\!\! \beta_0 +
  \case{48}{11}\beta_0^2 + \case5{11}\beta_0^3 \nonumber \\ &&\!\!\!\!
  - (\beta_0 + \case4{11}\beta_0^2 + \case5{11}\beta_0^3 ) \frac{
    \xi_{\rm d}(s;0) }{ \xi_{\rm q}(s;0) } \,,
\end{eqnarray}
\begin{eqnarray}
  \Delta_{\phi2}(s) \!\!\!\!&=&\!\!\!\! \case{10}3 \beta_0 +
  \case{16}{11}\beta_0^2 + \case{10}{77}\beta_0^3 \nonumber \\ 
  &&\!\!\!\! + (\case23 \beta_0 - \case4{33}\beta_0^2 -
  \case{10}{77}\beta_0^3 ) \frac{ \xi_{\rm d}(s;0) }{ \xi_{\rm q}(s;0)
    } \,,
\end{eqnarray}
\begin{eqnarray}\label{a5}
  \Delta_{\phi3}(s) \!\!\!\!&=&\!\!\!\! 3 + \case{93}{35}\beta_0 +
  \case{37}{105}\beta_0^2 + \case1{49}\beta_0^3 \nonumber \\ 
  &&\!\!\!\! + (1 + \case{27}{35}\beta_0 - \case1{105}\beta_0^2 -
  \case1{49}\beta_0^3 ) \frac{ \xi_{\rm d}(s;0) }{ \xi_{\rm q}(s;0) }
  \,,
\end{eqnarray}
\begin{equation}\label{a6}
  \Delta_{\psi1}(s) = \case{40}{11}\beta_0^2 + 2(\beta_0 +
  \case{13}{11}\beta_0^2) \frac{ \xi_{\rm d}(s;0) }{ \xi_{\rm q}(s;0)
    } \,,
\end{equation}
\begin{equation}
  \Delta_{\psi2}(s) = \case{40}{11}(\beta_0 + \case13\beta_0^2) +
  2(1 + \case{50}{33}\beta_0 + \case{13}{11}\beta_0^2) \frac{
    \xi_{\rm d}(s;0) }{ \xi_{\rm q}(s;0) } \,,
\end{equation}
\begin{eqnarray}\label{a8}
  \Delta_{\psi3}(s) \!\!\!\!&=&\!\!\!\! \beta_0^{-1} + 5 +
  \case{25}7\beta_0 + \case13\beta_0^2 \nonumber \\ &&\!\!\!\! +
  (\beta_0^{-1} + \case{19}7 + \case{61}{35}\beta_0 +
  \case{19}{105}\beta_0^2 ) \frac{ \xi_{\rm d}(s;0) }{ \xi_{\rm
      q}(s;0) } \,.
\end{eqnarray}

%======================================================================
\section*{References}

\rf{Alcock, C., \& Paczy\'nski, B. 1979, Nature, 281, 358}
\rf{Ballinger, W.E., Peacock, J.A., \& Heavens, A.F. 1996, MNRAS, 282,
  877}
\rf{Bardeen, J.M., Bond, J.R., Kaiser, N., \& Szalay, A.S. 1985, ApJ,
  304, 15}
\rf{Cole, S., Fisher, K.B., \& Weinberg, D.H. 1994, MNRAS, 267, 785}
\rf{Cole, S., Fisher, K.B., \& Weinberg, D.H. 1995, MNRAS, 275, 515}
\rf{Davis, M. \& Peebles, P.J.E. 1983, ApJ, 267, 465}
\rf{Fukugita, M., Shimasaku, K., \& Ichikawa, T. 1995, PASP, 107, 945}
\rf{Gunn, J.E., \& Weinberg, D.H. 1995, in Wide Field Spectroscopy and
  the Distant Universe, eds. S.J. Maddox \& A. Ara\'gon-Salamanca
  (World Scientific, Singapore), 3}
\rf{Hamilton, A.J.S. 1992, ApJ, 385, L5}
\rf{Kaiser, N. 1987, MNRAS, 227, 1}
\rf{Lahav, O, Lilje, P.B., Primack, J.R., \& Rees, M.J. 1991, MNRAS,
  251, 128}
\rf{Loveday, J., Peterson, B.A., Efstathiou, G., \& Maddox, S.J. 1992,
  ApJ, 390, 338}
\rf{Loveday, J., Efstathiou, G., Maddox, S.J., \& Peterson, B.A. 1996,
  ApJ, 468, 1}
\rf{Matsubara, T., \& Suto, Y. 1996, 470, L1}
\rf{Peacock, J.A., \& Dodds, S.J. 1994, MNRAS, 267, 1020}
\rf{Peebles, P.J.E. 1980, The Large-Scale Structure of the Universe
  (Princeton: Princeton Univ. Press)}
\rf{Peebles, P.J.E. 1993, Principles of Physical Cosmology (Princeton:
  Princeton Univ. Press)}
\rf{Ratcliffe, A., et al. 1996, MNRAS, 281, L47}
\rf{Ryden, B. 1995, ApJ, 452, 25}
\rf{Shectman, S.A., et al. 1996, ApJ, 470, 172}
\rf{Strauss, M.A. \& Willick, J. 1995, Phys. Rep., 261, 271}
\end{document}